
\voffset 0.1in
\hoffset 0.1in

\def\sumint{\hbox{$\sum$}\!\!\!\!\!\!\!\int}

\def\gboxit#1{\hbox{\vrule\vbox{\hrule\kern3pt\vtop
{\hbox{\kern3pt#1\kern3pt}
\kern3pt\hrule}}\vrule}}

\def\rr{\right\rangle}

\def\ttilde#1{\raise2ex\hbox{${\scriptscriptstyle(}\!
\sim\scriptscriptstyle{)}$}\mkern-16.5mu #1}
\def\dddots#1{\raise1ex\hbox{$^{\ldots}$}\mkern-16.5mu #1}
\def\siton#1#2{\raise1.5ex\hbox{$\scriptscriptstyle{#2}$}\mkern-16.5mu
#1}
\def\upleftarrow#1{\raise1.5ex\hbox{$\leftarrow$}\mkern-16.5mu #1}
\def\uprightarrow#1{\raise1.5ex\hbox{$\rightarrow$}\mkern-16.5mu #1}
\def\upleftrightarrow#1{\raise1.5ex\hbox{$\leftrightarrow$}\mkern-16.5mu
#1}
\def\bx#1#2{\vcenter{\hrule \hbox{\vrule height #2in \kern
#1in\vrule}\hrule}}

\def\squiggle#1{\lower1.5ex\hbox{$\sim$}\mkern-14mu #1}

\def\narrower{\advance\leftskip by\parindent \advance\rightskip
by\parindent}

\def\mbox#1#2{\vcenter{\hrule width#1in\hbox{\vrule height#2in
   \hskip#1in\vrule height#2in}\hrule width#1in}}
\def\eqsquare #1:#2:{\vcenter{\hrule width#1\hbox{\vrule height#2
   \hskip#1\vrule height#2}\hrule width#1}}
\def\inbox#1#2#3{\vcenter to #2in{\vfil\hbox to
#1in{$$\hfil#3\hfil$$}\vfil}}
\def\strutdepth{\dp\strutbox}
\def\marbul{\strut\vadjust{\kern-\strutdepth\specialbul}}
\def\specialbul{\vtop to \strutdepth{
    \baselineskip\strutdepth\vss\llap{$\bullet$\qquad}\null}}
\def\Bcomma{\lower6pt\hbox{$,$}}    
\def\bcomma{\lower3pt\hbox{$,$}}    

\def\sl{\scrsf}

\def\updots{\mathinner{\mskip 1mu\raise 1pt\hbox{.}
    \mskip 2mu\raise 4pt\hbox{.}\mskip 2mu
    \raise 7pt\vbox{\kern 7pt\hbox{.}}\mskip 1mu}}

\def\square{\kern1pt\vbox{\hrule height 1.2pt\hbox{\vrule width
1.2pt\hskip 3pt
   \vbox{\vskip 6pt}\hskip 3pt\vrule width 0.6pt}\hrule height
0.6pt}\kern1pt}
\def\ssquare{\kern1pt\vbox{\hrule height .6pt\hbox{\vrule width
.6pt\hskip 3pt
   \vbox{\vskip 6pt}\hskip 3pt\vrule width 0.6pt}\hrule height
0.6pt}\kern1pt}
\def\lege{\hbox{$ {     \lower.40ex\hbox{$>$}
		   \atop \raise.20ex\hbox{$<$}
		   }     $}  }

\def\rege{\hbox{$ {     \lower.40ex\hbox{$<$}
		   \atop \raise.20ex\hbox{$>$}
		   }     $}  }

\def\lapp{\hbox{$ {     \lower.40ex\hbox{$<$}
		   \atop \raise.20ex\hbox{$\sim$}
		   }     $}  }
\def\rapp{\hbox{$ {     \lower.40ex\hbox{$>$}
		   \atop \raise.20ex\hbox{$\sim$}
		   }     $}  }

\def\tridots{\hbox{$ {     \lower.40ex\hbox{$.$}
		   \atop \raise.20ex\hbox{$.\,.$}
		   }     $}  }
\def\Times{\times\hskip-2.3pt{\raise.25ex\hbox{{$\scriptscriptstyle|$}}}}

\def\rightonleft{\hbox{$ {     \lower.40ex\hbox{$\longrightarrow$}
		   \atop \raise.20ex\hbox{$\longleftarrow$}
		   }     $}  }

\def\pmb#1{\setbox0=\hbox{#1}%
\kern-.025em\copy0\kern-\wd0
\kern.05em\copy0\kern-\wd0
\kern-.025em\raise.0433em\box0 }
%
%
\font\fivebf=cmbx5
\font\sixbf=cmbx6
\font\sevenbf=cmbx7
\font\eightbf=cmbx8
\font\ninebf=cmbx9
\font\tenbf=cmbx10

\font\bfmone=cmbx10 scaled\magstep1

\font\sevenit=cmti7
\font\eightit=cmti8
\font\nineit=cmti9
\font\tenit=cmti10

\font\itmone=cmti10 scaled\magstep1

\font\fiverm=cmr5
\font\sixrm=cmr6
\font\sevenrm=cmr7
\font\eightrm=cmr8
\font\ninerm=cmr9
\font\tenrm=cmr10

\font\rmmone=cmr10 scaled\magstep1

\def\fontone{\def\rm{\fcm0\rmmone}%
  \textfont0=\rmmone \scriptfont0=\tenrm \scriptscriptfont0=\sevenrm
  \textfont1=\itmone \scriptfont1=\tenit \scriptscriptfont1=\sevenit
  \def\it{\fcm\itfcm\itmone}%
  \textfont\itfcm=\itmone
  \def\bf{\fcm\bffcm\bfmone}%
  \textfont\bffcm=\bfmone \scriptfont\bffcm=\tenbf
   \scriptscriptfont\bffcm=\sevenbf
  \tt \ttglue=.5em plus.25em minus.15em
  \normalbaselineskip=25pt
  \let\sc=\tenrm
  \let\big=\tenbig
  \setbox\strutbox=\hbox{\vrule height10.2pt depth4.2pt width\z@}%
  \normalbaselines\rm}



\font\ninerm=cmr9
\font\eightrm=cmr8
\font\sixrm=cmr6

\font\ninei=cmmi9
\font\eighti=cmmi8
\font\sixi=cmmi6
\skewchar\ninei='177 \skewchar\eighti='177 \skewchar\sixi='177

\font\ninesy=cmsy9
\font\eightsy=cmsy8
\font\sixsy=cmsy6
\skewchar\ninesy='60 \skewchar\eightsy='60 \skewchar\sixsy='60

\font\ninebf=cmbx9
\font\eightbf=cmbx8
\font\sixbf=cmbx6

\font\ninett=cmtt9
\font\eighttt=cmtt8

\hyphenchar\tentt=-1 
\hyphenchar\ninett=-1
\hyphenchar\eighttt=-1

\font\ninesl=cmsl9
\font\eightsl=cmsl8

\font\nineit=cmti9
\font\eightit=cmti8


\newskip\ttglue
\def\tenpoint{\def\rm{\fcm0\tenrm}%
  \textfont0=\tenrm \scriptfont0=\sevenrm \scriptscriptfont0=\fiverm
  \textfont1=\teni \scriptfont1=\seveni \scriptscriptfont1=\fivei
  \textfont2=\tensy \scriptfont2=\sevensy \scriptscriptfont2=\fivesy
  \textfont3=\tenex \scriptfont3=\tenex \scriptscriptfont3=\tenex
  \def\it{\fcm\itfcm\tenit}%
  \textfont\itfcm=\tenit
  \def\sl{\fcm\slfcm\tensl}%
  \textfont\slfcm=\tensl
  \def\bf{\fcm\bffcm\tenbf}%
  \textfont\bffcm=\tenbf \scriptfont\bffcm=\sevenbf
   \scriptscriptfont\bffcm=\fivebf
  \def\tt{\fcm\ttfcm\tentt}%
  \textfont\ttfcm=\tentt
  \tt \ttglue=.5em plus.25em minus.15em
  \normalbaselineskip=16pt
  \let\sc=\eightrm
  \let\big=\tenbig
  \setbox\strutbox=\hbox{\vrule height8.5pt depth3.5pt width\z@}%
  \normalbaselines\rm}

\def\ninepoint{\def\rm{\fcm0\ninerm}%
  \textfont0=\ninerm \scriptfont0=\sixrm \scriptscriptfont0=\fiverm
  \textfont1=\ninei \scriptfont1=\sixi \scriptscriptfont1=\fivei
  \textfont2=\ninesy \scriptfont2=\sixsy \scriptscriptfont2=\fivesy
  \textfont3=\tenex \scriptfont3=\tenex \scriptscriptfont3=\tenex
  \def\it{\fcm\itfcm\nineit}%
  \textfont\itfcm=\nineit
  \def\sl{\fcm\slfcm\ninesl}%
  \textfont\slfcm=\ninesl
  \def\bf{\fcm\bffcm\ninebf}%
  \textfont\bffcm=\ninebf \scriptfont\bffcm=\sixbf
   \scriptscriptfont\bffcm=\fivebf
  \def\tt{\fcm\ttfcm\ninett}%
  \textfont\ttfcm=\ninett
  \tt \ttglue=.5em plus.25em minus.15em
  \normalbaselineskip=11pt
  \let\sc=\sevenrm
  \let\big=\ninebig
  \setbox\strutbox=\hbox{\vrule height8pt depth3pt width\z@}%
  \normalbaselines\rm}

\def\eightpoint{\def\rm{\fcm0\eightrm}%
  \textfont0=\eightrm \scriptfont0=\sixrm \scriptscriptfont0=\fiverm
  \textfont1=\eighti \scriptfont1=\sixi \scriptscriptfont1=\fivei
  \textfont2=\eightsy \scriptfont2=\sixsy \scriptscriptfont2=\fivesy
  \textfont3=\tenex \scriptfont3=\tenex \scriptscriptfont3=\tenex
  \def\it{\fcm\itfcm\eightit}%
  \textfont\itfcm=\eightit
  \def\sl{\fcm\slfcm\eightsl}%
  \textfont\slfcm=\eightsl
  \def\bf{\fcm\bffcm\eightbf}%
  \textfont\bffcm=\eightbf \scriptfont\bffcm=\sixbf
   \scriptscriptfont\bffcm=\fivebf
  \def\tt{\fcm\ttfcm\eighttt}%
  \textfont\ttfcm=\eighttt
  \tt \ttglue=.5em plus.25em minus.15em
  \normalbaselineskip=9pt
  \let\sc=\sixrm
  \let\big=\eightbig
  \setbox\strutbox=\hbox{\vrule height7pt depth2pt width\z@}%
  \normalbaselines\rm}



\magnification=1200
\newbox\leftpage
\newdimen\fullhsize
\newdimen\hstitle
\newdimen\hsbody
\tolerance=600\hfuzz=2pt
\hoffset=0.1truein \voffset=0.1truein
\hsbody=\hsize \hstitle=\hsize
\font\titlefont=cmr10 scaled\magstep3
\font\secfont=cmbx10 scaled\magstep1

\def\nolabels{\def\eqnlabel##1{}\def\eqlabel##1{}\def\reflabel##1{}}
\def\writelabels{\def\eqnlabel##1{%
{\escapechar=` \hfill\rlap{\hskip.09in\string##1}}}%
\def\eqlabel##1{{\escapechar=` \rlap{\hskip.09in\string##1}}}%
\def\reflabel##1{\noexpand\llap{\string\string\string##1\hskip.31in}}}
\nolabels
\def\title#1{ \nopagenumbers\hsize=\hsbody%
\centerline{ {\titlefont #1} }%
\pageno=0}
\def\author#1{\vskip 1 truecm%
\centerline{{\sl #1}}%
\centerline{Center for Theoretical Physics}%
\centerline{Laboratory for Nuclear Science}%
\centerline{and Department of Physics}%
\centerline{Massachusetts Institute of Technology}%
\centerline{Cambridge, Massachusetts 02139}}
\def\abstract#1{\centerline{\bf ABSTRACT}\nobreak\medskip\nobreak\par #1}

\global\newcount\secno \global\secno=0
\global\newcount\meqno \global\meqno=1
\def\newsec#1{\global\advance\secno by1
\xdef\secsym{\ifcase\secno
\or I\or II\or III\or IV\or V\or VI\or VII\or VIII\or IX\or X\fi }
\global\meqno=1
\bigbreak\bigskip
\noindent{\secfont\secsym. #1}\par\nobreak\medskip\nobreak}
\xdef\secsym{}


\def\eqnn#1{\xdef #1{(\the\secno.\the\meqno)}%
\global\advance\meqno by1\eqnlabel#1}
\def\eqna#1{\xdef #1##1{\hbox{$(\the\secno.\the\meqno##1)$}}%
\global\advance\meqno by1\eqnlabel{#1$\{\}$}}
\def\eqn#1#2{\xdef #1{(\the\secno.\the\meqno)}\global\advance\meqno by1%
$$#2\eqno#1\eqlabel#1$$}
%
%
%

\def\eqnla#1#2#3{\xdef #2{(\the\secno.\the\meqno a)}
\xdef #1{(\the\secno.\the\meqno)}
$$#3\eqno#2\eqlabel#1\eqlabel#2$$}
\def\eqnlb#1#2{\xdef #1{(\the\secno.\the\meqno b)}
$$#2\eqno#1\eqlabel#1$$}
\def\eqnlc#1#2{\xdef #1{(\the\secno.\the\meqno c)}
$$#2\eqno#1\eqlabel#1$$}
\def\eqnld#1#2{\xdef #1{(\the\secno.\the\meqno d)}
$$#2\eqno#1\eqlabel#1$$}
\def\eqnle#1#2{\xdef #1{(\the\secno.\the\meqno e)}
$$#2\eqno#1\eqlabel#1$$}
\def\eqnlf#1#2{\xdef #1{(\the\secno.\the\meqno f)}
$$#2\eqno#1\eqlabel#1$$}
\def\eqnlbend#1#2{\xdef #1{(\the\secno.\the\meqno b)}\global\advance\meqno by1%
$$#2\eqno#1\eqlabel#1$$}
\def\eqnlcend#1#2{\xdef #1{(\the\secno.\the\meqno c)}\global\advance\meqno by1%
$$#2\eqno#1\eqlabel#1$$}
\def\eqnldend#1#2{\xdef #1{(\the\secno.\the\meqno d)}\global\advance\meqno by1%
$$#2\eqno#1\eqlabel#1$$}
\def\eqnleend#1#2{\xdef #1{(\the\secno.\the\meqno e)}\global\advance\meqno by1%
$$#2\eqno#1\eqlabel#1$$}
\def\eqnlfend#1#2{\xdef #1{(\the\secno.\the\meqno f)}\global\advance\meqno by1%
$$#2\eqno#1\eqlabel#1$$}
\def\eqnlgend#1#2{\xdef #1{(\the\secno.\the\meqno g)}\global\advance\meqno by1%
$$#2\eqno#1\eqlabel#1$$}
\global\newcount\ftno \global\ftno=1
\def\foot#1{{\baselineskip=12pt plus 1pt\footnote{$^{\the\ftno}$}{#1}}%
\global\advance\ftno by1}
\global\newcount\refno \global\refno=1
\newwrite\rfile
\def\ref#1#2{\the\refno\nref#1{#2}}
\def\nref#1#2{\xdef#1{\the\refno}%
\ifnum\refno=1\immediate\openout\rfile=refs.tmp\fi%
\immediate\write\rfile{\noexpand\item{#1\ }\reflabel{#1}#2}%
\global\advance\refno by1}
\def\addref#1{\immediate\write\rfile{\noexpand\item{}#1}}

\def\figures{\centerline{{\bf Figure Captions}}\medskip\parindent=40pt}
\def\fig#1#2{\medskip\item{Fig.~#1:  }#2}

\def\frac#1#2{{#1\over#2}}
\def\cm{{\cal M}}
\pageno=0
$^{\ref\mat{For recent reviews, see:
M. Dine, in proceedings of the Texas Symposium on Electroweak
Baryon Number Violation, New Haven, USA, 1992;
A.G. Cohen, D.B. Kaplan, and A.E. Nelson, UCSD 93-2/BU-HEP-93-4.}}$

$^{\ref\kili{D. Kirzhnits, JETP Lett. {\bf 15}, 529 (1972);
D. Kirzhnits and A. Linde, Phys. Lett. {\bf B 42}, 471
(1972);
D. Kirzhnits and A. Linde,
JETP {\bf 40}, 628 (1974).}}$
$^{\ref\doja{L. Dolan and R. Jackiw, Phys. Rev. {\bf D 9}, 3320 (1974).}}$
$^{\ref\wei{S. Weinberg, Phys. Rev. {\bf D 9}, 3357 (1974).}}$

$^{\ref\hsu{D.E. Brahm and S.D.H. Hsu, CALT-68-1705/HUTP-91-A063 (1991);
C.G. Boyd, D.E. Brahm and S.D.H. Hsu,
CALT-68-1795/HUTP-92-A027/EFI-92-22 (1992).}}$

$^{\ref\firef{M. Dine, R.G. Leigh, P. Huet, A. Linde, and D. Linde,
Phys. Rev. {\bf D 46}, 550 (1992) and Phys. Lett. {\bf B 283}, 319
(1992).}}$

$^{\ref\kapunp{J. Kapusta, D.B. Reiss, and S. Rudaz,
Nucl. Phys. {\bf B 263}, 207 (1986).}}$

$^{\ref\smref{M.E. Shaposhnikov, Phys. Lett. {\bf B 277}, 324 (1992);
Erratum-ibid. {\bf B 282}, 483 (1992).}}$

$^{\ref\car{M.E. Carrington, Phys. Rev. {\bf D 45}, 2933 (1992).}}$

$^{\ref\jai{V. Jain, Nucl. Phys. {\bf B 394}, 707 (1993).}}$

$^{\ref\zwi{J.R. Espinosa, M. Quiros, and F. Zwirner,
Phys. Lett. {\bf B 291}, 115 (1992), and CERN-TH-6577/92.}}$

$^{\ref\rengro{P. Elmfors, Int. J. Mod. Phys. {\bf A 8}, 1887 (1993)}}$

$^{\ref\taka{R. Takahashi, Z. Phys. C {\bf 26}, 601 (1985).}}$

$^{\ref\desyahm{W. Buchmueller and T. Helbig,
in proceedings of the International Warsaw Meeting on
Elementary Particle Physics, Kazimierz, Poland, 1992.}}$

$^{\ref\arn{P. Arnold, Phys. Rev. {\bf D 46}, 2628 (1992).}}$

$^{\ref\jaib{V. Jain,  MPI-Ph/92-72 (1992).}}$

$^{\ref\arndue{P. Arnold and O. Espinosa,
Phys. Rev. {\bf D 47}, 3546 (1993).}}$

$^{\ref\hebe{A. Hebecker, DESY-93-086 (1993).}}$

$^{\ref\gacpi{G. Amelino-Camelia and S.-Y. Pi,
Phys. Rev. {\bf D 47}, 2356 (1993).}}$

$^{\ref\hsulast{C.G. Boyd, D.E. Brahm and S.D.H. Hsu,
CALT-68-1858/HUTP-93-A011/EFI-93-22 (1992).}}$

$^{\ref\jmr{Investigations of finite temperature
phase transitions using renormalization group techniques
have been discussed, for example, in
C. Athorne and I.D. Lawrie, Nucl. Phys. {\bf B 265}, 552 (1986);
J. March-Russell, Phys. Lett. {\bf B 296}, 364 (1992);
F. Freire and C.R. Stephens, Z. Phys. {\bf  C 60}, 127 (1993).}}$

\vfill
\eject
\vfill

\centerline{{\bf SELF-CONSISTENTLY IMPROVED FINITE TEMPERATURE }}
\centerline{{\bf  EFFECTIVE
POTENTIAL FOR GAUGE THEORIES}\footnote{*}
{This work is
supported in part by funds provided by the U.S. Department of Energy (D.O.E.)
under contract $\#$DE-FG02-91ER40676. }}
\vskip 24pt
\centerline{to appear in Physical Review D49 (1994)}
\vskip 48pt
\bigskip

\centerline{G. Amelino-Camelia\footnote{**}
{Present Address: Center for Theoretical Physics,
Laboratory for Nuclear Science, and Department of Physics,
Massachusetts Institute of Technology, Cambridge, Massachusetts 02139 USA.
Email: amelino@mitlns.mit.edu.}}
\vskip 18pt
\baselineskip 12pt plus 0.2pt minus 0.2pt
\centerline{Department of Physics}
\centerline{Boston University}
\centerline{590 Commonwealth Ave.}
\centerline{Boston, Massachusetts 02215, USA}
\vskip 4.0cm

\vskip 1cm
\vfill
\noindent{BUHEP-93-12 \hfill}
\eject
\vfill
\baselineskip 24pt plus 0.2pt minus 0.2pt

\centerline{\bf ABSTRACT }
The finite temperature effective potential of the Abelian Higgs Model
is studied using the self-consistent composite operator method, which
can be used to sum up
the contributions of daisy and superdaisy diagrams.
The effect of the momentum dependence of the effective masses
is estimated
by using a Rayleigh-Ritz variational approximation.


\vfill
\eject

\newsec{Introduction}
It has been recently conjectured that the observed baryon asymmetry might
have been generated
at the electroweak scale if the phase transition is strongly
first order$^{[\mat]}$.
Unfortunately,
when the temperature $T$ is close
to the critical
temperature $T_c$
the finite temperature effective potential $V_T(\phi)$, which is an important
mathematical tool in the study of the phase transitions and can be used to
determine their order, cannot be evaluated reliably using the ordinary
perturbative approach$^{[\kili-\wei]}$;
in fact, at these
temperatures certain multi-loop diagrams become non-negligible even
when the coupling constants are very small.
In particular,
by using power counting
it has been argued$^{[\wei,\hsu]}$
that when $T \sim T_c$
there are important contributions from
the infinite classes of {\it daisy} and {\it superdaisy} diagrams
(represented in Fig.1),
which render
the ordinary one-loop approximation
of $V_T(\phi)$ unreliable
for all  $\phi < T$.
Therefore, the contributions from these diagrams must be added to the
one-loop result.
The corresponding improved approximation of $V_T(\phi)$
is expected$^{[\hsu,\firef]}$ to
be reliable (even when $T \sim T_c$) for
all $\phi > g T$ and up to order $g^3$, where $g$ is the
biggest coupling constant of the theory.

Due to the recent interest in the electroweak baryogenesis
several techniques of evaluation
of the contribution of daisy (and superdaisy) diagrams
to the finite temperature effective potential for various
theories have been presented in the literature$^{[\hsu-\jmr]}$.
In Ref.$[\gacpi]$,
in a study of the $\lambda \Phi^4$ scalar theory,
Pi and I have used
a method of resummation of the daisy and superdaisy diagrams
which
is based on
the generalization at finite temperature of the
Cornwall-Jackiw-Tomboulis$^{[\ref\corn{J.M. Cornwall,
R. Jackiw, and E. Tomboulis,
Phys. Rev. {\bf D 10}, 2428 (1974).}]}$
effective
potential for composite operators $V_{T}(\phi,G)$.
For bosonic quantum fields $\Phi(x)$, $V_{T}(\phi,G)$ is given by:
\eqn\vexp{V_T(\phi,G)=V_{cl}(\phi)
+{1 \over 2} TrLn D_0 G^{-1}
+{1 \over 2} Tr [ D^{-1} G - 1]
+ V_{T(2)}(\phi,G)  ~,}
\noindent
where$^{[\gacpi,\corn]}$ $G$ is a
possible expectation value of ${\cal T} \Phi(x) \Phi(y)$
(here ${\cal T}$ indicates time ordering),
$D_0$ is the free propagator, $D$ is the
tree-level propagator, $V_{cl}(\phi)$ is the classical potential,
and $V_{T(2)}(\phi,G)$ is given by
all the two-particle irreducible
vacuum-to-vacuum
graphs with two or more loops
in the theory with vertices given by the interaction part of the
shifted ($\Phi \rightarrow \Phi + \phi$) lagrangian and
propagators set equal to $G$.

The ordinary effective potential $V_{T}(\phi)$
is related to $V_{T}(\phi,G)$ by
the relation

\baselineskip 12pt plus 0.2pt minus 0.2pt
\eqnla\vtusu\vtusua{
V_T(\phi) = V_T(\phi,G_0)
{}~,}
\eqnlbend\vtusub{
\biggl[{\delta V_T(\phi,G) \over \delta G}\biggr]_{G=G_0}
= 0
{}~.}
\smallskip
\baselineskip 24pt plus 0.2pt minus 0.2pt

Using this formalism, it is easy
to see$^{[\gacpi,\hsulast]}$ that the resummation of the daisy and superdaisy
contributions corresponds to the following
approximation of the finite temperature effective potential:

\baselineskip 12pt plus 0.2pt minus 0.2pt
\eqnla\hf\hfa{
 V_T(\phi) \simeq V^{res}_T(\phi,G_0)
{}~,}
\eqnlb\hfb{
V^{res}_T(\phi,G) \equiv
V_{cl}(\phi)
+{1 \over 2} TrLn D_0 G^{-1}
+{1 \over 2} Tr [ D^{-1} G - 1]
+ \biggl[ V_{T(2)}(\phi,G) \biggr]_{O(g^2)}
{}~,}

\eqnlcend\hfc{
\biggl[{\delta V^{res}_T(\phi,G) \over \delta G}\biggr]_{G=G_0}
= 0
{}~,}
\smallskip
\baselineskip 24pt plus 0.2pt minus 0.2pt

\noindent
i.e. the expression for $V_T(\phi)$ given by \vexp\space and \vtusu\space
but with $V_{T(2)}(\phi,G)$ approximated by the leading
two-loop contributions in $G$.

The primary purpose of this paper is to address some technical issues
which often arise in the application of this
composite operator method, but are absent
in the leading order calculation for the $\lambda \Phi^4$ theory
presented in Ref.[\gacpi].

The most important technical complication  is the momentum dependence of
the effective masses (defined in the following section in analogy
with Ref.$[\gacpi]$), which appears when diagrams of the type in Fig.2b
are included in the approximation of $V_{T(2)}(\phi,G)$.
In the leading order calculation for the $\lambda \Phi^4$ scalar
theory such
contributions can be neglected. However, unless
one assumes that the gauge coupling is much smaller than the scalar
self-coupling, in the study of gauge theories
diagrams of the type in Fig.2b contribute in the leading order.
In the recent literature the importance of the problem of the momentum
dependence of the effective masses has been often pointed
out$^{[\zwi,\arndue,\hsulast]}$, but no consistent solution as been yet
developed. As a consequence, in most studies the finite temperature
effective potential has been calculated using the
{\it ad hoc} substitution of the effective masses by their value at
zero momentum.
I show that the structure of Eq.\hf,
in which the effective potential is
obtained as the solution of a variational problem,
naturally leads to a variational approximation which allows to
estimate analytically the effect of the momentum dependence of the
effective masses.
For definiteness and simplicity, I illustrate this technique
by studying the high temperature
effective potential of the Abelian Higgs Model,
which has been the subject
of numerous recent investigations$^{[\taka-\hebe]}$,
especially as a toy model of the Standard
Electroweak Model. My variational approximation of the daisy and
superdaisy resummed effective potential of the Abelian Higgs Model
agrees to order $e^3$ with the result of the ``improvement of the one-loop"
performed in Ref.[\car]. Moreover, I use the terms of order $e^4$ and
higher in my result to estimate how important the full higher order
correction can be expected to be,
and to show that an accurate evaluation of the effective potential
at the critical temperature also requires appropriate handling
of the momentum dependence of the effective masses.

The paper is organized as follows.
In Sec.II, I derive
for the Abelian Higgs Model the approximation of the finite temperature
effective potential for composite operators defined in Eq.\hfb, and the
{\it gap} equations \hfc.
Due to the momentum dependence of the effective masses,
the daisy and superdaisy resummed effective potential $V^{res}_T$
cannot be evaluated analytically.
In Sec.III, I make a
Rayleigh-Ritz variational approximation and evaluate
analytically the corresponding effective potential
in the high temperature limit.
Sec.IV is devoted to the discussion of the results and the conclusions.

\newsec{Daisy and Superdaisy Resummed Effective Potential}
In this section, using Eq.\hf, I study
the exact daisy and superdaisy resummed finite temperature effective
potential for the Abelian Higgs Model.
Here and in the following sections the analysis is performed in Landau
gauge.
Field theory at finite temperature T is
described using imaginary (Euclidean) time $\tau$ which is restricted to
the interval $0 \leq \tau \leq 1/T$.
The Feynman rules are the same as
those of ordinary field theory, except that the momentum space integral over
the time component
$k_4$ is replaced by a sum over discrete
frequencies $k_4 = i \pi n T$:
\eqn\disint{\int {d^4k \over (2 \pi)^4} \rightarrow T
\sum_n \int  {d^3k \over (2 \pi)^3} \equiv \sumint_k
{}~,}
where  $n$ is even (odd) for bosons (fermions).

The Lagrangian which describes the theory is
\eqn\lagori{\eqalign{
L=&-{1 \over 4} F_{\mu \nu} F^{\mu \nu}+
{1 \over 2} \partial_{\mu} \Phi_a \partial^{\mu} \Phi^a
+{1 \over 2} m^2 \Phi^2 - {\lambda \over 4!} \Phi^4 \cr
&-e \epsilon_{ab} \partial_{\mu} \Phi_a \Phi_b A^{\mu}+
{1 \over 2} e^2 \Phi^2 A^2
-{1 \over 2 \xi} (\partial_{\mu} A^{\mu})^2-
\eta^{+} \partial_{\mu} \partial^{\mu} \eta
{}~,}}
where
\eqn\abmunu{\eqalign{
&\Phi^2 \equiv \Phi_a \Phi^a~,~~~\Phi^4 \equiv (\Phi^2)^2 ~,~~~
a=1,2 ~,~~~b=1,2~,
\cr
&F_{\mu \nu} \equiv \partial_{\mu} A_{\nu} - \partial_{\nu} A_{\mu} ~,~~~
\mu=1,2,3,4 ~,~~~ \nu=1,2,3,4
{}~.}}

\noindent
$A_\mu$, $\Phi_a$, and $\eta$ are the gauge, Higgs, and ghost fields
respectively. Note that the ghost $\eta$ completely decouples from
the rest of the theory.
Because of the singularity of the inverse of the gauge boson propagator
in Landau gauge, it is appropriate to keep track of the gauge parameter
$\xi$, and take the limit $\xi \rightarrow 0$ only at the end.

After shifting the Higgs field $\Phi$
(i.e. $\Phi_1 \rightarrow \Phi_1 + \phi_1$ and
$\Phi_2  \rightarrow \Phi_2 + \phi_2$) from \lagori\space
one obtains
the following tree-level mass matrices
\eqn\mtre{\eqalign{
(m^2_\gamma)_{\mu \nu}&= e^2 \phi^2 g_{\mu \nu}
\cr
(m^2_\Phi)_{ab}&= (-m^2+{\lambda \over 6} \phi^2) \delta_{ab} +
{\lambda \over 3} \phi_a \phi_b
{}~}}
\noindent
for the gauge and the Higgs fields respectively,
and the
following classical potential
\eqn\vtcl{V_{cl}(\phi)=
-{1 \over 2} m^2 \phi^2 + {\lambda \over 4!} \phi^4
{}~.}
\noindent
The propagator in momentum space is given by
\eqn\protre{\eqalign{
(D^{-1}(\phi; k))_{\mu \nu}&=
(e^2 \phi^2 - k^2)
({k_{\mu} k_{\nu} \over k^2}
- g_{\mu \nu}  ) +
({k^2 \over \xi} - e^2 \phi^2) {k_{\mu} k_{\nu} \over k^2}
\cr
(D^{-1}(\phi ;k))_{ab}&= (m^2_\Phi)_{ab} - \delta_{ab} k^2
\cr
(D^{-1}(\phi ;k))_{a \mu}&= - i e k_{\mu} \epsilon_{ab} \phi_b
{}~.}}
\noindent
where $k^2 \equiv k_\mu k^\mu$.

In the application of the composite operator method,
it is first necessary to identify,
based on the magnitude of the coupling constants,
the diagrams which give the
leading contribution to $V_{T(2)}(\phi,G_0)$.
As argued in Ref.[\arndue],
in order to justify the high temperature approximation and
to have a well-defined loop expansion, one needs $e^4 << \lambda << e^2$;
however, the scenario is different depending on
whether $e^3 << \lambda << e^2$ or $e^4 << \lambda \le e^3$.
If $e^3 << \lambda << e^2$
$V_{T(2)}$ is approximated
by the sum of the four diagrams in Fig.3 whereas if $e^4 << \lambda \le e^3$
the diagram in Fig.3a is neglected.
I study the richer scenario $e^3 << \lambda << e^2$
(the results for the other case can be obtained as the appropriate
limit of the results for $e^3 << \lambda << e^2$);
in these hypotheses, $V^{res}_T$ can be written as

\baselineskip 12pt plus 0.2pt minus 0.2pt
\eqnla\hfahm\hfahma{\eqalign{
V^{res}_T =&
V_{cl}(\phi)
+{1 \over 2} TrLn D_0 G_0^{-1}
+{1 \over 2} Tr [ D^{-1} G_0 - 1]
\cr
&+ V_2^{(a)}(G_0)
+ V_2^{(b)}(G_0)
+ V_2^{(c)}(G_0)
+ V_2^{(d)}(G_0)
{}~,}}
\eqnlb\hfahmb{V_2^{(a)}(G) \equiv - {\lambda \over 4!} \sumint_p  \sumint_q
[G_{aa}(p) G_{bb}(q) + 2 G_{ab}(p) G_{ba}(q)]
{}~,}
\eqnlc\hfahmc{V_2^{(b)}(G) \equiv - {e^2 \over 2} g^{\mu \nu}
\sumint_p  \sumint_q
G_{\mu \nu}(q) G_{aa}(p)
{}~,}
\eqnld\hfahmd{V_2^{(c)}(G) \equiv {e^2 \over 4} \sumint_p  \sumint_q
\epsilon_{ab} \epsilon_{cd} (2p+q)^\mu (2p+q)^\nu
G_{\mu \nu}(q) G_{ad}(p) G_{bc}(p+q)
{}~,}
\eqnleend\hfahme{V_2^{(d)}(G) \equiv {e^2 \over 2} \sumint_p  \sumint_q
 \epsilon_{ac} \epsilon_{db} (2q+p)^\mu (2q+p)^\nu
G_{ab}(p+q) G_{c \nu}(p) G_{\mu d}(q)
{}~.}
\baselineskip 24pt plus 0.2pt minus 0.2pt

\noindent
where (see \hf) $G_0$ is the solution of
\eqn\statahm{{\delta V^{res}_T \over \delta G}=0 ~.}

Using \hfahm, \statahm\space can be rewritten as the following
system of gap equations:

\baselineskip 12pt plus 0.2pt minus 0.2pt
\eqnla\hfgapahm\hfgapahma{\eqalign{
G^{-1}_{ab}(k)  = & D^{-1}_{ab}(k)
- 2 {\delta V_2 \over \delta G_{ba}(k)} \cr
= & D^{-1}_{ab}(k) +
{\lambda \over 6} \sumint_p [2 G_{ab}(p) + \delta_{ab} G_{cc}(p)]
+ e^2
\delta_{ab} \sumint_p  g^{\mu \nu} G_{\mu \nu}(p) \cr
& + e^2 \sumint_p
\epsilon_{ad} \epsilon_{bc} (2k+p)^\mu (2k+p)^\nu
G_{\mu \nu}(p) G_{cd}(p+k)  \cr
&+ e^2 \sumint_p
\epsilon_{ad} \epsilon_{bc} (p+k)^\mu (2k-p)^\nu
G_{\mu d}(k-p) G_{c \nu}(p)
{}~,}}
\eqnlb\hfgapahmb{\eqalign{
G^{-1}_{\mu \nu}(k)  = & D^{-1}_{\mu \nu}(k)
- 2 {\delta V_2 \over \delta G_{\nu \mu}(k)} \cr
= & D^{-1}_{\mu \nu}(k) + e^2
\sumint_p  g_{\mu \nu} G_{aa}(p) \cr
& +  {e^2 \over 2} \sumint_p
\epsilon_{ab} \epsilon_{dc} (2p+k)_\mu (2p+k)_\nu
G_{ad}(p) G_{bc}(p+k)
{}~,}}
\eqnlcend\hfgapahmc{\eqalign{
G^{-1}_{a \mu}(k) = & D^{-1}_{a \mu}(k)
- 2 {\delta V_2 \over \delta G_{\mu a}(k)}
\cr
= & D^{-1}_{a \mu}(k) + e^2 \sumint_p
\epsilon_{ab} \epsilon_{cd} (2p+k)_\mu (2k+p)^\nu
G_{db}(p+k) G_{c \nu}(p)
{}~,}}
\baselineskip 24pt plus 0.2pt minus 0.2pt

\noindent
where $V_2 \equiv V_2^{(a)} +V_2^{(b)} +V_2^{(c)} +V_2^{(d)}$ .
The diagrammatic representation of the various $\delta V_2 / \delta G_{xy}$
is given in Fig.4.

In order to study the system of equations \hfgapahm\space I introduce
effective masses by taking an {\it ansatz} for the
propagator:

\baselineskip 12pt plus 0.2pt minus 0.2pt
\eqnla\ansex\ansexa{
G^{-1}_{\mu \nu}=
(\cm^2_t(k) - k^2) t_{\mu \nu}(k) +
(\cm^2_l(k) - k^2) l_{\mu \nu}(k) +
({k^2 \over \xi} - e^2 \phi^2) {k_{\mu} k_{\nu} \over k^2}
{}~,}
\eqnlb\ansexb{
G^{-1}_{ab}= (\cm_\Phi^2(k))_{ab} - \delta_{ab} k^2
{}~,}

\eqnlcend\ansexc{
G^{-1}_{a \mu}=
- i e k_{\mu} \epsilon_{ab} \phi_b
{}~,}
\baselineskip 24pt plus 0.2pt minus 0.2pt

\noindent
where $t_{\mu \nu}$ and $l_{\mu \nu}$ are defined by
(N.B. $\{\mu,\nu\}=1,2,3,4$; $\{i,j\}=1,2,3$; ${\bf k}^2 \equiv k_i k^i$)
\eqn\ptpl{\eqalign{
t_{\mu \nu}(k) \equiv &
\delta_{\mu i} \delta_{\nu j}  (\delta^{ij} - {k^i k^j \over {\bf k}^2})
\cr
l_{\mu \nu}(k) \equiv & {k_\mu k_\nu \over k^2} - g_{\mu \nu}  - t_{\mu \nu}
{}~.}}
The Eqs.\ansex\space express the propagator in terms of effective
masses $\cm_x(k)$
and are consistent in Landau gauge with the Eqs.\hfgapahm.
The two transverse modes of the
gauge boson acquire the same effective
mass $\cm_t(k)$ whereas the longitudinal mode has an independent
effective mass $\cm_l(k)$, as required by the way Lorentz invariance is
broken at finite temperature$^{[\ref\pisarsk{D.J. Gross, R.D. Pisarski, and
L.G. Yaffe, Rev. Mod. Phys. {\bf 53}, 43 (1981).}]}$.
Also note that the symmetry of Eq.\hfgapahm\space
under exchange of $\phi_1$ and $\phi_2$
(a consequence of the analogous property of the propagator $D$) indicates
that $(\cm_\Phi^2(k))_{12} = (\cm_\Phi^2(k))_{21}$.

Using \ansex\space and the following properties
of $t_{\mu \nu}$ and $l_{\mu \nu}$
\eqn\proptpl{\eqalign{
t_{\mu \sigma}(k) ~ t^{\sigma \nu}(k) =& - t^{~ \nu}_\mu(k) ~~,~~~
t_{\mu}^{~ \mu}(k) = -2 ~,
\cr
l_{\mu \sigma}(k) ~ l^{\sigma \nu}(k) =& - l^{~ \nu}_\mu(k) ~~,~~~
l_{\mu}^{~ \mu}(k) = -1
{}~,}}
the equations \hfgapahm\space can be rewritten in terms of
the effective masses as

\baselineskip 12pt plus 0.2pt minus 0.2pt
\eqnla\rompiballe\rompiballea{\eqalign{
(\cm_\Phi^2(k))_{ab}  = & (-m^2+{\lambda \over 6} \phi^2) \delta_{ab} +
{\lambda \over 3} \phi_a \phi_b
\cr
&+{\lambda \over 6} \sumint_p
{2 \epsilon_{ac} \epsilon_{bd} (\cm_\Phi^2(p))_{cd}
+ \delta_{ab} [(\cm_\Phi^2(p))_{cc}-4 p^2]
\over [(\cm_\Phi^2(p))_{11}-p^2] [(\cm_\Phi^2(p))_{22}-p^2]-
[(\cm_\Phi^2(p))_{12}]^2}
\cr
&- e^2
\delta_{ab} \sumint_p
\biggl( {2 \over \cm^2_t(p) - p^2}+
{1 \over \cm^2_l(p)-p^2} \biggr) \cr
& + e^2 \sumint_p
\epsilon_{ad} \epsilon_{bc} (2k+p)^\mu (2k+p)^\nu
\biggl( {t_{\mu \nu}(p)\over \cm^2_t(p) - p^2}+
{l_{\mu \nu}(p) \over \cm^2_l(p) - p^2} \biggr)
\cr
&~~~{\epsilon_{cm} \epsilon_{dn} (\cm_\Phi^2(p+k))_{mn}
- \delta_{cd} (p+k)^2
\over [(\cm_\Phi^2(p+k))_{11}-p^2] [(\cm_\Phi^2(p+k))_{22}-p^2]-
[(\cm_\Phi^2(p+k))_{12}]^2}
{}~,}}
\smallskip
\eqnlbend\rompiballeb{\eqalign{
\cm^2_s(k) = & e^2 \phi^2
- e^2
\sumint_p
{(\cm_\Phi^2(p))_{cc}- 2 p^2
\over [(\cm_\Phi^2(p))_{11}-p^2] [(\cm_\Phi^2(p))_{22}-p^2]-
[(\cm_\Phi^2(p))_{12}]^2}
\cr
& +  {(1+\rho_{s}) e^2 \over 4} \sumint_p
\epsilon_{ab} \epsilon_{dc} (2p+k)_\mu (2p+k)_\nu W_s^{\mu \nu}(k)
\cr
&~~~{\epsilon_{am} \epsilon_{dn} (\cm_\Phi^2(p))_{mn}
- \delta_{ad} p^2
\over [(\cm_\Phi^2(p))_{11}-p^2] [(\cm_\Phi^2(p))_{22}-p^2]-
[(\cm_\Phi^2(p))_{12}]^2}
\cr
&~~~{\epsilon_{bh} \epsilon_{ci} (\cm_\Phi^2(p+k))_{hi}
- \delta_{bc} (p+k)^2
\over [(\cm_\Phi^2(p+k))_{11}-p^2] [(\cm_\Phi^2(p+k))_{22}-p^2]-
[(\cm_\Phi^2(p+k))_{12}]^2}
{}~,}}
\baselineskip 24pt plus 0.2pt minus 0.2pt

\medskip
\noindent
where $s=t,l$, $\rho_t=0$, $\rho_l=1$, $W_t^{\mu \nu} \equiv t^{\mu \nu}$,
and $W_l^{\mu \nu} \equiv l^{\mu \nu}$.
In \rompiballe\space the Landau gauge limit $\xi \rightarrow 0$ has
been taken; no divergence occurs because of a cancellation
between the $1/\xi$
terms which appear in $G^{-1}_{\mu \nu}$ (as given in \ansexa)
and $D^{-1}_{\mu \nu}$.

The Eqs.\rompiballe\space show that,
because of the contributions corresponding to the non-local self-energy
diagrams in Fig.4c and 4g (the last terms on the r.h.s.
of Eqs.\rompiballea\space and \rompiballeb),
the effective masses must have a highly non-trivial dependence
on the momentum $k$, and
this makes the search for solutions very difficult.
[Note that in the leading order calculation for the $\lambda \Phi^4$
theory, no non-local
self-energy diagram
contributes to the gap equation for the effective mass$^{[\gacpi]}$.
As a result the gap equation can be solved analytically in the
high-temperature approximation.]

However,  $V_T^{res}$ could be evaluated from
Eqs.\hfahm\space
and \hfgapahm\space (after renormalization) using numerical methods.
The fact that
the exact daisy and superdaisy resummed effective
potential is given by Eqs.\hfahm\space and \hfgapahm\space is a first
important result of the composite operator method.

\newsec{Rayleigh-Ritz Approximation}
In this section I study $V_T^{res}$ analytically using
the observation that
an approximate solution of the variational
problem \hf\space can be obtained by
evaluating $V_T^{res}(\phi,G)$ with
specific parameter-dependent expressions for the
propagator $G(k)$ and then varying these
parameters.
This type of procedure is known$^{[\corn]}$
as the ``Rayleigh-Ritz variational approximation''.
I shall take as the parameter-dependent $G(k)$ the expression
\ansex\space with all the momentum dependent ``exact'' effective
masses $\cm_x(k)$ replaced by constant ``Rayleigh-Ritz effective
masses'' $M_x$. This consistently leads, as we will see,  to momentum
independent gap equations for the $M_x$.

The present approximation is quite different from the approximation
which has been frequently used in the recent literature, where one makes
the ad hoc assumption $\cm_x(k) \simeq \cm_x(0)$ in the gap equations
for the effective masses.
The differences are twofold. First, in my approximation the
self-consistency
between the effective potential and the
gap equations is preserved whereas in the ``$\cm_x(k) \simeq \cm_x(0)$
approximation'' this self-consistency is lost.
Second, as I shall argue in Sec.IV based on the analysis of the results
obtained in the following,
some average effect
of the  dependence of the diagrams
in Fig.4c and 4g on the external momentum is reflected in
my approximation of $V_T^{res}$, even though the Rayleigh-Ritz effective
masses are
constant.
In the $\cm_x(k) \simeq \cm_x(0)$ approximation this
momentum dependence is completely neglected, leading to an
uncontrolled (and probably large, as argued in Refs.[\hsu,\arn,\arndue]) error
in the evaluation of the daisy
and superdaisy diagrams. [The spirit of performing the approximation at
the level of the expectation value of the Hamiltonian rather than at the
level of the equations that follow from varying the exact Hamiltonian is
very similar to the difference between the Kohn-Sham approximation and
the Slater approximation in Hartree-Fock many-body theory: the former is
an approximation to the Hamiltonian expectation value, which is than
varied, the latter is an
approximation to the variational equations.
It is known that the
Kohn-Sham method is
better$^{[\ref\betjac{Discussion of this subject can be found in:
H.A. Bethe and R. Jackiw,
{\it Intermediate Quantum Mechanics}, third ed. (The Benjamin/Cummings
Publishing Company,
Menlo Park, California, 1986).}]}$.
This result should encourage the use of the Rayleigh-Ritz approximation;
however, because of the differences between the problem at hand
and Hartree-Fock many-body theory, it does not necessarily imply that
the $\cm_x(k) \simeq \cm_x(0)$ approximation
be inadequate.]

The number of independent Rayleigh-Ritz effective masses
to be varied can be reduced by using
the fact that the symmetry of the lagrangian suggests that
$V_T^{res}$ depends on $\phi^2 \equiv \phi_1^2+\phi_2^2$ rather than
separately on $\phi_1$ and $\phi_2$.
This
invariance of $V_T^{res}$ under rotations in $\phi$-space
allows
to perform the calculations starting from a diagonal tree-level
mass matrix for the Higgs $(m^2_\Phi)_{ab}=
(-m^2+{\lambda \over 2} \phi^2) \delta_{a1} \delta_{b1} +
(-m^2+{\lambda \over 6} \phi^2) \delta_{a2} \delta_{b2}$,
and in this hypothesis one can consistently assume that also the
Rayleigh-Ritz mass matrix for the Higgs is diagonal.
This completes the definition of the variational problem
which is studied in this section; the corresponding approximation
of the daisy and superdaisy resummed finite temperature effective potential
can be formally written as

\baselineskip 12pt plus 0.2pt minus 0.2pt
\eqnla\rr\rra{
V^{res}_T \simeq V^{res}_T(\phi,G(\{M_0\}))
{}~,}

\eqnlbend\rrb{
\biggl[{\delta V^{res}_T(\phi,G(\{M\} ))
\over \delta M^n}\biggr]_{\{ M \}=\{M_0 \}}
= 0
{}~.}
\smallskip
\baselineskip 24pt plus 0.2pt minus 0.2pt

\noindent
where $\{M \} \equiv \{M^1,M^2,M^3,M^4 \} \equiv
\{ M_\phi, M_\chi, M_t, M_l \}$,
$V^{res}_T(\phi,G)$ is defined in Eq.\hfahm,
and $G(\{M\} )$ is given by

\baselineskip 12pt plus 0.2pt minus 0.2pt
\eqnla\ansrr\ansrra{
G^{-1}_{\mu \nu}=
(M^2_t - k^2) t_{\mu \nu}(k) +
(M^2_l - k^2) l_{\mu \nu}(k) +
({k^2 \over \xi} - e^2 \phi^2) {k_{\mu} k_{\nu} \over k^2}
{}~,}
\eqnlb\ansrrb{
G^{-1}_{ab}= \delta_{a1} \delta_{b1} (M_\phi^2 - k^2)+
\delta_{a2} \delta_{b2} (M_\chi^2 - k^2)
{}~,}

\eqnlcend\ansrrc{
G^{-1}_{a \mu}= - i e k_{\mu} \epsilon_{ab} \phi_b
{}~.}
\baselineskip 24pt plus 0.2pt minus 0.2pt

The effective potential $V^{res}_T(\phi,G(\{M\} ))$
in Eq.\rr\space includes divergent integrals; therefore a
regularization and renormalization procedure is necessary.
The self-consistency of Eqs.\vexp-\vtusu\space
implies$^{[\corn]}$
that the effective potential $V_{T}(\phi,G)$ and the gap equations are
renormalizable.
Such a renormalization has been discussed in detail in the analog study
of the daisy and superdaisy resummed effective potential
of the $\lambda \Phi^4$ scalar theory presented in Ref.$[\gacpi]$.
In that case it has been shown that the only effect of renormalization
on the high temperature part of the effective potential is the substitution
of the bare parameters with renormalized ones.
This is due to the fact that the dominant high temperature
contributions to the effective potential come from the infrared and are
not sensitive to the ultraviolet
behavior$^{[\ref\notaben{Note that in selecting the diagrams to be
included in a given high temperature approximation one might neglect
some diagrams with important ultraviolet
contributions and small high temperature part. This could render
the approximation nonrenormalizable, but still the high temperature
part of the results should be reliable.}]}$.
In the following I shall assume that the same applies in the case of the
Abelian Higgs Model
and therefore, rather than performing renormalization
explicitly, I will simply omit the (zero-temperature) ultraviolet
contributions and
substitute in my high temperature effective potential the
bare parameters with renormalized ones.

The gap equations \rrb\space can be put in simple form by performing
the high temperature approximation
of $V^{res}_T(\phi,G(\{M\} ))$.
Using the well-known results$^{[\doja]}$
\baselineskip 24pt plus 0.2pt minus 0.2pt
\baselineskip 30pt plus 0.2pt minus 0.2pt
\eqn\tadp{\eqalign{
& {1 \over 2} TrLn[k^2-y^2]  \equiv {1 \over 2} \sumint_k
\ln[-(n \pi T)^2-{\bf k}^2-y^2]
\simeq -{ \pi^2 T^4 \over 90} + {y^2 T^2 \over 24}
- {y^3 T \over 12 \pi}
+ {c_\Omega y^4 \over 32 \pi^2} \cr
& \Omega(y)  \equiv \sumint_k {1 \over (n \pi T)^2+{\bf k}^2+y^2 }
={1 \over y} {\partial \over \partial y} \biggl\{
{1 \over 2} TrLn[k^2-y^2] \biggr\}
\simeq {T^2 \over 12} - {T y \over 4 \pi} + {c_\Omega \over 8 \pi^2} y^2 \cr
& ~~~~~~~~~~~~~~~~~~~~ c_\Omega \equiv {1 \over 2} \ln({T^2 \over \sigma^2})
+{1 \over 2}+ \ln(4 \pi) - \gamma_{Euler}
{}~,}}
\baselineskip 24pt plus 0.2pt minus 0.2pt

\noindent
(where $\sigma$ is a renormalization scale and as planned
the (zero-temperature) ultraviolet divergent contributions have been
omitted), the high-temperature
approximations of $(TrLnG^{-1})/2$ and $(Tr[D^{-1}G -1])/2$ in Landau
gauge are easily obtained:

\baselineskip 12pt plus 0.2pt minus 0.2pt
\eqnla\restv\restva{\eqalign{
{1 \over 2}TrLnG^{-1} \simeq &
{1 \over 2} TrLn[(k^2-M_\phi^2) (k^2-M_\chi^2) (k^2-M_t^2)^2 (k^2-M_l^2)]
\cr
\simeq & -{\pi^2 T^4 \over 18}
+{ T^2  \over 24} (M^2_\phi+M^2_\chi+ 2 M^2_t+M^2_l) \cr
&-{ T  \over 12 \pi} (M^3_\phi+M^3_\chi+ 2 M^3_t+M^3_l)
\cr
&+{ c_\Omega  \over 32 \pi^2} (M^4_\phi+M^4_\chi+ 2 M^4_t+M^4_l)
{}~,}}

\eqnlbend\restvb{\eqalign{
{1 \over 2}Tr[D^{-1}G -1]=& {1 \over 2}Tr[(D^{-1} - G^{-1})G ]
\cr
= &{1 \over 2} (m^2_\phi-M^2_\phi) \Omega(M_\phi)+
{1 \over 2} (m^2_\chi-M^2_\chi) \Omega(M_\chi) \cr
& +(e^2 \phi^2 - M^2_t) \Omega(M_t)
+ {1 \over 2} (e^2 \phi^2 - M^2_l) \Omega(M_l) \cr
\simeq&(m^2_\phi-M^2_\phi)
({T^2 \over 24} - {T M_\phi \over 8 \pi}
+{ c_\Omega  \over 16 \pi^2} M^2_\phi) \cr
&+(m^2_\chi - M^2_\chi)
({T^2 \over 24} - {T M_\chi \over 8 \pi}
+{ c_\Omega  \over 16 \pi^2} M^2_\chi) \cr
&+2(e^2 \phi^2-M^2_t)({T^2 \over 24} - {T M_t \over 8 \pi}
+{ c_\Omega  \over 16 \pi^2} M^2_t) \cr
&+(e^2 \phi^2 - M^2_l) ({T^2 \over 24} - {T M_l  \over 8 \pi}
+{ c_\Omega  \over 16 \pi^2} M^2_l)
{}~,}}
\smallskip
\baselineskip 24pt plus 0.2pt minus 0.2pt

\noindent
where $m^2_{\phi} \equiv -m^2 + {\lambda \phi^2}/2$ and
$m^2_{\chi} \equiv -m^2 + {\lambda \phi^2}/6$.
Note that in \restva\space I neglected an unimportant infinite constant
(i.e. independent of the effective masses) which appears as a
consequence of the $1/\xi$ pole in the determinant
of the $G^{-1}_{\mu \nu}$ of Eq.\ansrra.
In obtaining \restvb\space the limit $\xi \rightarrow 0$
presented no complication because of the cancellation
between the $1/\xi$
terms which appear in $D^{-1}_{\mu \nu} - G^{-1}_{\mu \nu}$
(see \protre\space
and \ansrr).
In the following evaluations of $V_2^{(a)}$, $V_2^{(b)}$,
$V_2^{(c)}$, and $V_2^{(d)}$ the limit $\xi \rightarrow 0$ is also taken
and no divergences appear.

$V_2^{(a)}(G(\{M \}))$ and $V_2^{(b)}(G(\{M \}))$ have the same
structure as the two-loop contribution which is discussed in detail
in Ref.$[\gacpi]$; their high-temperature approximations
are given by:

\baselineskip 12pt plus 0.2pt minus 0.2pt
\eqnla\vii\viia{\eqalign{
V_2^{(a)}&(G(\{M \})) = {\lambda \over 24} \biggl[
3 \Omega^2(M_\phi) + 3 \Omega^2(M_\chi)
+ 2 \Omega(M_\phi) \Omega(M_\chi) \biggr] \cr
\simeq & {\lambda T^2 \over 432}
- {\lambda T^3 (M_\phi+M_\chi) \over 144 \pi}
+ {\lambda T^2 M_\phi M_\chi \over 192 \pi^2}
+ {\lambda T^2 (M^2_\phi+M^2_\chi) \over  \pi^2}
({c_\Omega \over 288} + {1 \over 128})
{}~,}}

\eqnlbend\viib{\eqalign{
V_2^{(b)}&(G(\{M \})) = {e^2 \over 2} \biggl(
2 \Omega(M_t) + \Omega(M_l) \biggr)
\biggl( \Omega(M_\phi) + \Omega(M_\chi) \biggr)
\cr
\simeq & {e^2 T^2 \over 48}
- {e^2 T^3 (4 M_t + 2 M_l + 3 M_\phi + 3 M_\chi) \over 96 \pi}
+ {e^2 T^2 (2 M_t + M_l) (M_\phi + M_\chi) \over 32 \pi^2} \cr
& + {e^2 T^2 c_\Omega
(4 M_t^2 + 2 M_l^2 + 3 M^2_\phi + 3 M^2_\chi) \over  192 \pi^2}
{}~.}}
\smallskip
\baselineskip 24pt plus 0.2pt minus 0.2pt

The corresponding approximation
of $V_2^{(c)}(G(\{M \}))$, based on results
obtained in Refs.$[\arndue,\ref\parwani{R.R. Parwani,
Phys. Rev. {\bf D 45}, 4695 (1992).}]$,
is discussed in the appendix; the result is
\eqn\viic{\eqalign{
V_2^{(c)}&(G(\{M \})) \simeq - {e^2 T^2 \over 288}
+ {e^2 T^3 (2 M_t - M_l) \over 48 \pi}  \cr
&+ {e^2 T^2 (M_t^2 -2 M^2_\phi - 2 M^2_\chi) \over  32 \pi^2}
\ln({M_t + M_\phi + M_\chi \over 3 T})
\cr
& + {e^2 T^2 (M_\phi M_\chi - M_t M_\phi - M_t M_\chi) \over 32 \pi^2}
+ {e^2 T^2
( c_{\Theta t}+ c_{\Theta l}) (M_\phi^2 + M_\chi^2) \over  64 \pi^2}
\cr
& - {e^2 T^2 [(3 c_{\Theta t}+8 c_\Omega) M_t^2 +
(3 c_{\Theta l} -4 c_\Omega) M_l^2] \over  384 \pi^2}
{}~.}}
$c_{\Theta t}$ and $c_{\Theta l}$ are defined in the appendix.

Finally, the contribution to $V_T^{res}$ from the diagram in Fig.3d
can be easily calculated
by noticing that Eq.\ansrr\space implies
that in Landau gauge $G_{a \mu}=0$ :
\eqn\vbd{V_2^{(d)}(G(\{M \}))=0
{}~.}
[Note that \vbd\space
is consistent with Eqs.\protre, \hfgapahmc, and \ansrr.]

Using \restv, \vii, \viic, and \vbd\space it
is now possible to express the gap
equations \rrb\space in the following form:

\baselineskip 12pt plus 0.2pt minus 0.2pt
\eqnla\gapm\gapma{\eqalign{M^2_{\phi(\chi)} \simeq & m^2_{\phi(\chi)}
+({ \lambda \over 18} + {e^2 \over 4}) T^2
-\biggl[a
-{1 \over \pi} \ln ({M_t + M_\phi + M_\chi \over 3 T}) \biggr]
e^2 T M_{\phi(\chi)} \cr
& -({ \lambda \over 24 \pi}
+ {e^2 \over 4 \pi}) T M_{\chi(\phi)}
-{e^2 \over 4 \pi} T (M_t + M_l) \cr
&+ {e^2 T \over 4 \pi}
{ 2 M^2_{\phi} + 2  M_{\chi}^2 - M^2_t \over
M_{\phi} + M_{\chi} + M_t}
+{ c_\Omega  \over  \pi} { M^3_{\phi(\chi)}  \over T}
{}~,}}
\eqnlb\gapmb{\eqalign{M^2_t \simeq & e^2 \phi^2
+\biggl[
{c_{\Theta t} \over 16 \pi}
-{1 \over 4 \pi} \ln ({M_t + M_\phi + M_\chi \over 3 T}) \biggr]
e^2 T M_t \cr
& - {e^2 T \over 8 \pi}
\biggl[ M_{\phi} + M_{\chi} +
{ M_t^2 - 2 M^2_{\phi} - 2  M_{\chi}^2
\over
M_{\phi} + M_{\chi} + M_t}
\biggr]
+{ c_\Omega  \over  \pi} { M^3_t  \over T}
{}~,}}
\eqnlcend\gapmc{\eqalign{M^2_l \simeq e^2 \phi^2 + {e^2 \over 3} T^2
- e^2 ({c_\Omega \over 3 \pi} - {c_{\Theta l} \over 8 \pi}) T M_{l}
- {e^2 T \over 4 \pi} (M_{\phi} + M_{\chi})
+{ c_\Omega  \over  \pi} { M^3_l  \over T}
{}~,}}
\baselineskip 24pt plus 0.2pt minus 0.2pt

\noindent
where $a \equiv
[(c_\Omega + c_{\Theta t} + c_{\Theta l} )/(4 \pi)]
+\lambda  [(4 c_\Omega + 9)/(72 \pi e^2)]$.

Note that in these calculations one can assume
that $ e T < \phi < T$
because, as already
discussed in the introduction, the daisy and
superdaisy resummed effective potential is expected to give a meaningful
approximation of the full effective
potential only for $\phi > e T $, and
the daisy and
superdaisy diagrams should be negligible when $\phi>T$.
This observation combined with
the Eqs.\gapm\space justifies {\it a posteriori} the high
temperature approximations which I performed; in fact,
the effective masses in \gapm\space are of order $eT$ or $e \phi$,
so that $M_{\phi, \chi, t, l}/T<1$ when
$e T < \phi < T$.

The Eqs.\gapm\space also indicate that
$M_t < M_{\phi, \chi, l}$
because only \gapmb\space does not contain a contribution of order $e^2 T^2$;
moreover, \gapma\space indicates that
$M^2_{\phi} \simeq M^2_{\chi}$ (they differ only
at order $\lambda \phi^2$).
Therefore one can make the following approximations
\eqn\appro{\eqalign{
\ln ({M_\phi + M_\chi + M_t \over 3 T}) \simeq &
\ln ({M_\phi + M_\chi \over 3 T}) + {M_t \over M_{\phi}+M_{\chi}} ~,
\cr
{2 M_{\phi}^2 + 2 M_{\chi}^2 - M_t^2
\over M_{\phi}+M_{\chi}+M_t} \simeq & M_{\phi}+M_{\chi}-M_t
{}~,}}
which allow to rewrite the gap equations \gapm\space as

\baselineskip 12pt plus 0.2pt minus 0.2pt
\eqnla\appgapm\appgapma{\eqalign{M^2_{\phi(\chi)} \simeq & m^2_{\phi(\chi)}
+({ \lambda \over 18} + {e^2 \over 4}) T^2
-\biggl[a-{1 \over 4 \pi}
-{1 \over \pi} \ln ({M_\phi + M_\chi \over 3 T}) \biggr]
e^2 T M_{\phi(\chi)} \cr
& - { \lambda \over 24 \pi} T M_{\chi(\phi)}
- {e^2 \over 4 \pi} T M_l
+{ c_\Omega  \over  \pi} { M^3_{\phi(\chi)}  \over T}
{}~,}}
\eqnlb\appgapmb{M^2_t \simeq  e^2 \phi^2
+\biggl[
{c_{\Theta t} \over 16 \pi}
-{1 \over 8 \pi}
-{1 \over 4 \pi} \ln ({M_\phi + M_\chi \over 3 T}) \biggr]
e^2 T M_t +{ c_\Omega  \over  \pi} { M^3_t  \over T}
{}~,}
\eqnlcend\appgapmc{M^2_l \simeq e^2 \phi^2 + {e^2 \over 3} T^2
- e^2 ({c_\Omega \over 3 \pi} - {c_{\Theta l} \over 8 \pi}) T M_{l}
- {e^2 T \over 4 \pi} (M_{\phi} + M_{\chi})
+{ c_\Omega  \over  \pi} { M^3_l  \over T}
.}
\baselineskip 24pt plus 0.2pt minus 0.2pt

Using \vtcl, \restv, \vii, \viic,
\vbd, \appro, and \appgapm\space one finds that the
desired
Rayleigh-Ritz
and high-temperature approximation of the daisy and superdaisy resummed
finite temperature effective potential for the
Abelian Higgs Model is given by
(omitting unimportant $\phi$-independent contributions):
\eqn\vresult{\eqalign{
V^{res}_T(\phi,\{ M_0 \}) \simeq &
-{1 \over 2} m^2 \phi^2 + {\lambda \over 4!} \phi^4
+{ T^2  \over 24} (m^2_\phi+m^2_\chi+ 3 e^2 \phi^2) \cr
&-{ T  \over 12 \pi} (M^3_{\phi,0}+M^3_{\chi,0}+ 2 M^3_{t,0}+M^3_{l,0})
\cr
&+ {e^2 T^2 (2 M^2_{\phi,0} + 2 M^2_{\chi,0} - M_{t,0}^2) \over  32 \pi^2}
\ln({M_{\phi,0} + M_{\chi,0} \over 3 T})
\cr
&- {e^2 T^2  \over 32 \pi^2} M_{l,0} (M_{\phi,0}+M_{\chi,0})
+  ({e^2 \over 32} - {\lambda \over 192})  { T^2  \over  \pi^2}
M_{\phi,0} M_{\chi,0}
\cr
&- {e^2 T^2  \over  \pi^2}
({c_{\Omega}  \over 48 } + {c_{\Theta t}  \over 128}) M_{l,0}^2
+\tilde{a} {e^2 T^2  \over  \pi^2} (M^2_{\phi,0}+M^2_{\chi,0})
\cr
&+{c_{\Theta t}  \over 128 \pi^2} e^2 T^2 M_{t,0}^2
+{3 c_\Omega  \over 32 \pi^2}
(M^4_{\phi,0}+M^4_{\chi,0}+ 2 M^4_{t,0}+M^4_{l,0})
,}}
where $M_{\phi,0}$, $M_{\chi,0}$, $M_{t,0}$, and $M_{l,0}$ are the solutions
of the gap equations \appgapm, and
\space\space\space\space\space\space\space\space\space\space\space\space
$\tilde{a} \equiv 1/32 - (c_{\Theta t}+c_{\Theta l})/64
+(\lambda / e^2)(c_\Omega /288 - 1/128)$.

Note that in \vresult\space all terms linear in the effective masses
have cancelled out$^{[\ref\notejs{A similar cancellation has been found
in Ref.[\gacpi] for the effective potential of the $\lambda \Phi^4$
scalar theory.}]}$.
In the literature there has been an extensive debate
on the possibility
that the resummation of the daisy and superdaisy diagrams
might induce contributions to the finite temperature
effective potential which are linear in the effective masses.
Using the general form of the Rayleigh-Ritz approximation
with momentum independent effective masses one can
show$^{[\ref\gacbanf{G. Amelino-Camelia, to appear in proceedings
of the III Workshop on Thermal Field Theories and Their Applications,
Banff, Canada, August 16-27, 1993.}]}$
that a cancellation of the linear terms always
occurs.

The Rayleigh-Ritz variational approximation
and the high-temperature expansion have led to a
great simplification of the very
complicated expression of $V_T^{res}$ obtained in the equations
\hfahm\space and \hfgapahm.
Still, as given by \appgapm\space and \vresult, $V_T^{res}$ cannot be
calculated analytically, but now the required numerical
evaluations are very simple.
A first estimate of
the effective potential in \vresult\space
can be obtained analytically by using the following
approximations:
$M_{\phi,0} \simeq M_{\chi,0}$,
$\ln (M^2_{\phi,0}/T^2) \simeq \ln(\lambda/18+e^2/4)$,
$M_{\phi,0} \rightarrow (e^2 T^2/4 + \lambda T^2/18)^{1/2}$
in \appgapmc,
and $M_{l,0} \rightarrow (e^2 T^2/3)^{1/2}$ in \appgapma.
These approximations
are reliable when $\phi << T$ (see \appgapm).

\newsec{Discussion and Conclusions}
As discussed in the introduction, the daisy and superdaisy improved
effective potential $V_T^{res}(\phi)$ should give a reliable
approximation
of the full effective  potential $V_T(\phi)$
in the high temperature limit, but when
$T \sim T_c$ one expects $V_T^{res}(\phi) \simeq V_T(\phi)$ only
for $\phi > e T_c$.
This implies that the evaluation of $V_T^{res}(\phi)$  is sufficient
in the investigation of strongly first order phase transitions (if
at the critical temperature the asymmetric minimum $\phi_c$ is
greater than $e T_c$, $V_T^{res}(\phi)$ should reliably
determine
the existence and the position of $\phi_c$),
but $V_T^{res}(\phi)$ cannot be used to discriminate between a
second order and a very weekly first order phase transition.
Recent models of baryogenesis at the electroweak phase transition
require that this transition be strongly first order, and therefore
the evaluation of $V_T^{res}(\phi)$ for the Standard Electroweak Model
should lead to a reliable test of these models.

Also note that, because I neglected the contributions to
$V_{T(2)}$ from diagrams of order $e^4$ and higher (see Fig.5),
the result \vresult\space gives a reliable approximation of the full
effective potential up to order $e^3$.
The terms in \vresult\space of
order $e^4$ and higher can be used to
estimate how important the complete higher order correction
should be expected
to be, and to verify whether or not the differences between the
Rayleigh-Ritz approximation performed in the
preceding section and
the $\cm_x(k) \simeq \cm_x(0)$ approximation used in
the literature can result in relevantly different physical predictions.

The comparison of the
Rayleigh-Ritz approximation with
the $\cm_x(k) \simeq \cm_x(0)$ approximation is indeed the next
topic that I want to comment on.
Besides the conceptual issues which I already discussed, probably the
clearest shortcoming of the $\cm_x(k) \simeq \cm_x(0)$ approximation
is that it completely ignores the contribution of the self-energy
diagram in Fig.4c; in fact, this
diagram vanishes in the zero external momentum limit.
In the Rayleigh-Ritz approximation the contributions to the gap
equations which correspond to the diagram in Fig.4c are the ones coming
from $\delta V_2^{(c)} / \delta M_{\phi (\chi)}$;
Eqs.\viic\space and \gapma\space show that these contributions are
certainly non-negligible, and, in particular, for
small $e$ the term of
order $e^2 T M_\phi \ln(M_{\phi (\chi)}/T)
\sim e^2 \ln(e) T M_{\phi(\chi)}$
is the dominant term linear in
$M_{\phi (\chi)}$ in \gapma.
Even for the self-energy diagram in Fig.4g, which
does not vanish in the
zero external momentum limit, there are significant quantitative
differences between the Rayleigh-Ritz approximation and the
$\cm_x(k) \simeq \cm_x(0)$ approximation.
Clearly the Rayleigh-Ritz approximation
accounts for some average effect
of the dependence on the external momentum of
the diagrams in Fig.4c and 4g,
which is the origin of the momentum dependence of the
exact effective masses $\cm_x(k)$.

These differences in the structure of the gap equations between the
Rayleigh-Ritz approximation and the
$\cm_x(k) \simeq \cm_x(0)$ approximation result in important
quantitative differences at the level of the effective  potential.
In Fig.6 it is shown that the two approximations lead to relevantly
different predictions for the shape of the effective potential
at the critical temperature; in particular the position of the asymmetric
minimum $\phi_c$ differs by 20-30$\%$.
Some physical phenomena at the phase transitions depend critically
on the value of $\phi_c$ (for example the rate of baryon
number violation at the electroweak phase transition is exponentially
sensitive to $\phi_c$) and therefore Fig.6 shows that an accurate
analysis of these phenomena requires a proper handling  of the momentum
dependence of the effective masses.

Concerning the nature of the phase transition of the Abelian Higgs Model
it is useful to
notice (again using $e T << \phi << T$)
that the equations \appgapm\space and \vresult\space imply
that: (I) besides the expected contributions involving even
powers of $\phi$, there
is a negative contribution of order $e^3 T \phi^3$
to the effective potential, which
comes from the $T M_t^3$ term,
and (II) there is no contribution
of order $e^3 T^3 \phi$.
These observations indicate$^{[\hsu,\firef]}$ that
there is a critical temperature $T_c$ at which $V_T^{res}(\phi)$
has two degenerate minima.
{}From Eq.\vresult\space it is also easy to realize that $\phi_c >e T_c$
when $e^2/\lambda>>1$, which indicates that at least
in these hypotheses the Abelian Higgs Model has a first order phase
transition. This is the same qualitative conclusion that one
reaches using the one-loop improved effective potential evaluated in
Ref.[\car]; indeed it is easy to verify
that Eq.\vresult\space agrees with the improved one-loop result to order
$e^3$.
However, it is important to observe that at the critical temperature
the complete $V_T^{res}$, including the contributions of order $e^4$
and higher, is relevantly different from the one-loop improved effective
potential (see Fig.7); in particular the
position of the asymmetric minimum receives a significant correction.
This should suggest some caution concerning the accuracy
of recent predictions which have been obtained using
the improved one-loop approximation, like the lower limit on the
Higgs mass
for successful baryogenesis at the one Higgs doublet electroweak
phase transition.

I conclude by emphasizing that
the techniques discussed in this analysis of the
Abelian Higgs Model clearly
apply to any gauge theory.
Apart from the obvious complication
of having to deal with a richer particle content,
even in the study of more complex gauge theories
the major elements of
difficulty will still be: (I)
the momentum dependence of the effective masses which is introduced
by non-local self-energy diagrams of the type generally represented in
Fig.2a, and (II) the evaluation of diagrams of the type
generally represented in Fig.2b.
Using the composite operator method,
one can do better than the daisy and superdaisy resummation
by going beyond the present lowest order in the
coupling approximation of $V_{T(2)}$.
Also my Rayleigh-Ritz approximation can be
improved (so that the effect of the momentum dependence of the exact
effective masses $\cm_x(k)$ is estimated even more accurately) by using
more elaborated versions of the parameter dependent expression for $G$;
for example, one can make the
substitutions
$M_{x}^2 \rightarrow  M_{x}^2 + Y_{x} {\bf k}^2$
in Eq.\ansrr\space
and vary not only the $M_{x}$'s but also the additional parameters $Y_{x}$.

\bigskip

\bigskip

\bigskip
\centerline{{\bf ACKNOWLEDGEMENTS}}

I am most grateful to S.-Y. Pi for sharing with me unpublished results
of her investigations of this problem and for many discussions, especially
suggesting the Rayleigh-Ritz approximation.
I acknowledge partial support from the ``Fondazioni
Angelo Della
Riccia'', Firenze, Italy.

\vfill
\eject
\vfill
\def\appendix#1#2{\global\meqno=1\xdef\secsym{\hbox{#1}}\bigbreak\bigskip
\noindent{\bf Appendix #1. #2}\par\nobreak\medskip\nobreak}
\def\eqn#1#2{\xdef #1{(A.\the\meqno)}\global\advance\meqno by1%
$$#2\eqno#1\eqlabel#1$$}
\appendix{ } {$~~$High Temperature approximation of $V_2^{(c)}$}

This Appendix is devoted to the
high temperature approximation of $V_2^{(c)}(G(\{ M \})) \equiv$ \space
\space \space \space \space \space \space \space \space \space
$V_2^{(c)}(M_\phi,M_\chi,M_t,M_l)$, which
is used in the calculations of Sec.III.
As it can be inferred
from Sec.II and III, $V_2^{(c)}(M_\phi,M_\chi,M_t,M_l)$ is given by
\eqn\vbc{\eqalign{
V_2^{(c)}(M_\phi,M_\chi,M_t,M_l) = & {e^2 \over 2} \sumint_p \sumint_q
{(2p+q)^\mu (2p+q)^\nu \over [M_\phi^2 - p^2] [M_\chi^2 - (p+q)^2]}
\biggl( {t_{\mu \nu}(q)\over M_t^2 - q^2}+
{l_{\mu \nu}(q) \over M_l^2 - q^2} \biggr)
{}~. }}

The high temperature approximation
of $V_2^{(c)}(M_\phi,M_\chi,M_t,M_l)$ has been
evaluated in Ref.$[\arndue]$ in the limit $M_t=M_l$.
The result of $[\arndue]$ is based on the observation that
\eqn\vomethe{\eqalign{
V_2^{(c)}(M_\phi,M_\chi,M_\gamma,M_\gamma) =
- {e^2 \over 2} \biggl\{ &
\Omega(M_\gamma) [\Omega(M_\phi)+\Omega(M_\chi)]
- \Omega(M_\phi) \Omega(M_\chi)
\cr
& +(M_\gamma^2 - 2 M^2_\phi - 2 M^2_\chi) \Theta(M_\phi,M_\chi,M_\gamma)
\cr
& + {M^2_\phi -  M^2_\chi \over M_\gamma^2}
[\Omega(M_\gamma) - \Omega(0)] [\Omega(M_\phi) - \Omega(M_\chi)]
\cr
& + {(M^2_\phi -  M^2_\chi)^2 \over M_\gamma^2}
[\Theta(M_\phi,M_\chi,M_\gamma) - \Theta(M_\phi,M_\chi,0)]
\biggr\}
{}~, }}
where $\Omega(y)$ has been defined in \tadp, and
\eqn\teta{\eqalign{
\Theta(x,y,z) \equiv &  \sumint_p \sumint_q
{1 \over [x^2 - q^2]  [y^2 - p^2] [z^2 - (p+q)^2]}
{}~. }}
Eq.\vomethe, combined with the high temperature approximation
of $\Theta(x,y,z)$ which was obtained in Refs.$[\arndue,\parwani]$
and the corresponding approximation of $\Omega(y)$ which
is reported in \tadp,
leads to the high temperature
approximation of $V_2^{(c)}(M_\phi,M_\chi,M_\gamma,M_\gamma)$.
For the analysis of Sec.III it is sufficient to consider the limit
$(M_\phi-M_\chi)^2 << (M_\phi+M_\chi)^2$; in this hypothesis one finds
\eqn\varndue{\eqalign{
V_2^{(c)}(M_\phi,M_\chi,M_\gamma,M_\gamma) \simeq e^2 \biggl\{ &
-{T^4 \over 288} + {T^3 M_\gamma \over 48 \pi}
+ {T^2 \over 32 \pi^2} (M_\phi  M_\chi - M_\gamma M_\phi - M_\gamma M_\chi)
\cr
& + {T^2 \over 32 \pi^2} (M_\gamma^2 - 2 M^2_\phi - 2 M_\chi^2)
\ln\biggl( {M_\gamma + M_\phi + M_\chi \over 3 T} \biggr)
\cr
& -\biggl({c_\Omega \over 96 \pi^2} + {c_\Theta \over 128 \pi^2} \biggr)
T^2 M_\gamma^2
+ {c_\Theta \over 64 \pi^2}
T^2 (M^2_\phi + M^2_\chi)
\biggr\}
{}~, }}
where the $\phi$-independent quantity $c_\Theta$ is analog to
the $c_\Omega$ of Eq.\tadp, and is expressed
in Ref.$[\parwani]$ as a combination of
integrals which can be evaluated numerically.
[Note that in \varndue, following the strategy outlined in Sec.III, I
omitted some ultraviolet divergent contributions.]

The calculation of $V_2^{(c)}(M_\phi,M_\chi,M_t,M_l)$ is more difficult
than the one of
$V_2^{(c)}(M_\phi,M_\chi,$ $M_\gamma,M_\gamma)$
because only when $M_t=M_l$
the integrand in \vbc\space takes the simple form which leads to the
relation \vomethe.
Rather than proceeding to its
explicit evaluation, I shall obtain the high temperature
approximation of $V_2^{(c)}(M_\phi,M_\chi,M_t,M_l)$ by using as a guide
the result \varndue\space and by exploiting
the fact that $V_2^{(c)}(M_\phi,M_\chi,M_\gamma,0)$ and
$V_2^{(c)}(M_\phi,M_\chi,0,M_\gamma)$ have very different analytic properties.
[N.B. Eq.\vbc\space implies that
$V_2^{(c)}(M_\phi,M_\chi,M_t,M_l)=V_2^{(c)}(M_\phi,M_\chi,M_t,0)+
V_2^{(c)}(M_\phi,M_\chi,0,M_l)$.]

The first important observation is that terms of the
type $T^2 M_\gamma M_{\phi(\chi)}$ can only be
present in $V_2^{(c)}(M_\phi,M_\chi,M_\gamma,0)$.
In fact, this type of nonanalytic
term$^{[\ref\nonan{The presence in the high temperature expansion
of terms of the type $T^2 M_{\gamma} M_{\phi(\chi)}$ would not be possible
if $V_2^{(c)}$ were an analytic function of the squares of the effective
masses. A good introduction to the
relation between infrared problems of some finite temperature diagrams,
nonanalyticity, and the $k_4=0$ term of the sum over the discretized
fourth component of the momenta can be found in: J. Kapusta,
{\it Finite Temperature Field Theory} (Cambridge University Press,
Cambridge, England, 1989).}]}$
originates from the peculiar infrared properties of the diagram in Fig.3c
and it
gets a contribution only from the $p_4=0$, $q_4=0$ term of the sums in
\vbc. By looking at
the structure of $l_{\mu \nu}$ one
sees that $V_2^{(c)}(M_\phi,M_\chi,0,M_\gamma)$ could not include such
contributions.

Similarly, it is easy to realize that terms of the type of the one in
Eq.\varndue\space which is
proportional to $\ln(M_\gamma+M_\phi+M_\chi)$ (which is also
nonanalytic)
can only contribute to $V_2^{(c)}(M_\phi,M_\chi,M_\gamma,0)$. This is best
seen by tracing back the calculation of Refs.$[\arndue,\parwani]$ and
noticing that also this term comes from the part of the gauge boson
propagator which is proportional to $t_{\mu \nu}$.

Finally, one can notice that the terms $e^2 T^3 M_\gamma / 48 \pi$ and
$-e^2 c_\Omega T^2 M^2_\gamma / 96 \pi^2$ in \varndue\space originate from
the following contribution to $V_2^{(c)}(M_\phi,M_\chi,M_\gamma,M_\gamma)$
\eqn\vbc{\eqalign{
- {e^2 \over 2}  \sumint_q    {1 \over M_\gamma^2 - q^2} &
\biggl[ \sumint_p
4 {p^2-{(p q)^2 \over q^2} \over [M_\phi^2 - p^2] [M_\chi^2 - (p+q)^2]}
\biggr]_{q=0;M_{\phi(\chi)}=0} \simeq \cr
& \simeq e^2 \biggl[ - {T^4 \over 144 }
+ {T^3 M_\gamma \over 48 \pi}
- {c_\Omega T^2 M^2_\gamma \over 96 \pi^2} \biggr]
{}~. }}
In the expression of $V_2^{(c)}(M_\phi,M_\chi,M_t,M_l)$ the contribution
\vbc\space should be substituted by
\eqn\vbc{\eqalign{
- {e^2 \over 2} \biggl\{ \sumint_q            &
{1 \over M_t^2 - q^2}
\biggl[ \sumint_p
{4 p^\mu p^\nu t_{\mu \nu}(q)
\over [M_\phi^2 - p^2] [M_\chi^2 - (p+q)^2]}
\biggr]_{q=0;M_{\phi(\chi)}=0}
\cr
\sumint_q            &
{1 \over M_l^2 - q^2}
\biggl[ \sumint_p
{4 p^\mu p^\nu l_{\mu \nu}(q)
\over [M_\phi^2 - p^2] [M_\chi^2 - (p+q)^2]}
\biggr]_{q=0;M_{\phi(\chi)}=0} \biggr\} \simeq \cr
& \simeq e^2 \biggl[ - {T^4 \over 144 }
+ {T^3 (2 M_t - M_l) \over 48 \pi}
- {c_\Omega T^2 (2 M^2_t - M^2_l) \over 96 \pi^2} \biggr]
{}~. }}

These observations, combined with \varndue, lead to the conclusion that
\eqn\viicapp{\eqalign{
V_2^{(c)}&(M_\phi,M_\chi,M_t,M_l) \simeq - {e^2 T^2 \over 288}
+ {e^2 T^3 (2 M_t - M_l) \over 48 \pi}  \cr
&+ {e^2 T^2 (M_t^2 -2 M^2_\phi - 2 M^2_\chi) \over  32 \pi^2}
\ln({M_t + M_\phi + M_\chi \over 3 T})
\cr
& + {e^2 T^2 (M_\phi M_\chi - M_t M_\phi - M_t M_\chi) \over 32 \pi^2}
+ {e^2 T^2
( c_{\Theta t}+ c_{\Theta l}) (M_\phi^2 + M_\chi^2) \over  64 \pi^2}
\cr
& - {e^2 T^2 [(3 c_{\Theta t}+8 c_\Omega) M_t^2 +
(3 c_{\Theta l} -4 c_\Omega) M_l^2] \over  384 \pi^2}
{}~,}}
where the $\phi$-independent quantities
$c_{\Theta t}$ and $c_{\Theta l}$
verify $c_{\Theta t}+c_{\Theta l}=c_{\Theta}$ and, like
$c_\Theta$, are given by
combinations of
integrals which can be evaluated numerically$^{[\parwani]}$.

\vfill
\eject
\vfill
\vfill\eject\immediate\closeout\rfile
\centerline{{\bf References}}\bigskip{
\catcode`\@=11\escapechar=` %
\input refs.tmp\vfill\eject}
\vfill
\eject
\vfill
\figures
\fig1 Examples of (a,b) daisy and (c,d) superdaisy diagrams.
\vskip0.4cm
\fig2 When diagrams of shape (b) are included in the approximation of
$V_{T(2)}(\phi,G)$, $\delta V^{res}_T(\phi,G) / \delta G$
contains contributions from self-energy diagrams of shape (a), which
lead to momentum dependent effective masses.
\vskip0.4cm
\fig3 The diagrams of order (a) $\lambda$ or (b,c,d) $e^2$ which
contribute to $V_{T(2)}$.
\vskip0.4cm
\fig4 The diagrammatic representation of the integrals which contribute
to (a,b,c,d) $\delta V_2 / \delta G_{ab}$,
(e) $\delta V_2 / \delta G_{a \mu}$, and
(f,g) $\delta V_2 / \delta G_{\mu \nu}$.
\vskip0.4cm
\fig5 Examples of diagrams which are neglected in the present lowest order
in the couplings approximation of $V_{T(2)}$.
\vskip0.4cm
\fig6 The plot at $T=T_c$,  $e=.2$, and $\lambda=.01$ of
$\Delta v(\phi) \equiv
10^7 Real[V_T(\phi)-V_T(0)]/T_c^4$ versus $\phi/e T_c$ for the
Rayleigh-Ritz approximation (solid curve) and for the
$\cm_x(k) \simeq \cm_x(0)$ approximation (dot curve).
\vskip0.4cm
\fig7 The plot at $T=T_c$,  $e=.2$, and $\lambda=.01$
of $\Delta v(\phi) \equiv
10^7 Real[V_T(\phi)-V_T(0)]/T_c^4$ versus $\phi/e T_c$
for the Rayleigh-Ritz approximation
of the effective potential, described in Eq.\vresult, (solid curve)
and for the effective potential which is obtained by neglecting
all the contributions of order $e^4$ and higher
in Eq.\vresult\space (dot curve).

\vfill
\end

\bigskip

\bigskip

\bigskip

\bigskip

\bigskip

$^{[\gacpi]}$

---------------------

$^{[\ref\fen{P.
Fendley, Phys. Lett. {\bf B 196}, 175 (1987).}]}$

----------------------

\eqn\p{
{}~,}

\eqn\pp{\eqalign{
{}~,}}

--------------------------

and two-loop ($V^{II}$) contributions.

\baselineskip 12pt plus 0.2pt minus 0.2pt
\eqnla\s\sa{
{}~,}
\eqnlb\sb{
{}~,}
\eqnlc\sc{
{}~,}
\eqnld\sd{
{}~,}
where
\eqnle\se{
{}~,}
and now $G(x,x)$ is given by
\eqnlfend\sf{
{}~.}
\baselineskip 24pt plus 0.2pt minus 0.2pt

\noindent
and the interaction Lagrangian $L_{int}$ is at least cubic in $\Phi$.

----------------------
------------------------

\bigskip

\bigskip

\bigskip

\bigskip

\bigskip

\bigskip

$^{[\gacpi]}$

---------------------

$^{[\ref\fen{P.
Fendley, Phys. Lett. {\bf B 196}, 175 (1987).}]}$

----------------------

\eqn\p{
{}~,}

\eqn\pp{\eqalign{
{}~,}}

--------------------------

and two-loop ($V^{II}$) contributions.

\baselineskip 12pt plus 0.2pt minus 0.2pt
\eqnla\s\sa{
{}~,}
\eqnlb\sb{
{}~,}
\eqnlc\sc{
{}~,}
\eqnld\sd{
{}~,}
where
\eqnle\se{
{}~,}
and now $G(x,x)$ is given by
\eqnlfend\sf{
{}~.}
\baselineskip 24pt plus 0.2pt minus 0.2pt

\noindent
and the interaction Lagrangian $L_{int}$ is at least cubic in $\Phi$.

----------------------
------------------------

\bigskip

---------------------

sources $J$ and $K$ defined by

\baselineskip 12pt plus 0.2pt minus 0.2pt
\eqnla\cjt\cjta{\eqalign{Z  \equiv &  I
\cr
& K(x,y)
.}}
\baselineskip 24pt plus 0.2pt minus 0.2pt

\noindent
[We have set $c=\hbar=1$.]
oundary conditions the classical Euclidean action, which may be written as

\baselineskip 12pt plus 0.2pt minus 0.2pt
\eqnlb\cjtb{
, ~}

\noindent
where $D_0(x-y)$  is the free propagator

\eqnlcend\cjtc{
{}~,}
\baselineskip 24pt plus 0.2pt minus 0.2pt

\noindent
and the interaction Lagrangian $L_{int}$ is at least cubic in $\Phi$.

**************************************************************************
In particular, the Rayleigh-Ritz approximation
should be very useful in studying analytically
the high temperature effective potential of
phenomenologically relevant
theories.
**************************************************************************
$^{[\ref\dinelin{Using the language of ordinary perturbation
theory, Refs.$[\firef,\zwi]$ also argued that such linear terms
should not be present.}]}$.
**************************************************************************
; therefore, also in the study
of the Abelian Higgs Model the technique of approximation
proposed in $[\gacpi]$ gives a powerful method of
exact resummation of the daisy and superdaisy contributions to
the finite temperature effective potential.
This conclusion also applies to any other gauge theory; in fact,
it is easy to realize that both in
\hfahm\space and \hfgapahm\space
and in the analogous equations for other gauge theories
the only major new elements of difficulty (compared to the
study of the $\lambda \Phi^4$)
are the necessary momentum dependence of the effective masses
and the evaluation
of diagrams of the type in Fig.2c and 2d.

**************************************************************************

\noindent
As a check of the meaningfulness of the performed approximations, it is
interesting to notice that for $\phi=0$ Eq.\appgapmb\space admits
the solution $M_t=0$; this is what
one would expect as a consequence of gauge
invariance$^{[\firef,\pisarsk]}$.

**************************************************************************
in fact, using
the gap equations \appgapm\space one can rewrite
\vresult\space as
\eqn\vmlin{\eqalign{
V^{res}_T&(\phi,\{ M \}) \simeq
{1 \over 2} m^2 \phi^2 + {\lambda \over 4!} \phi^4
+{ T^2  \over 24} (m^2_\phi+m^2_\chi+ 3 e^2 \phi^2)
-{ T  \over 12 \pi} (M^3_\phi+M^3_\chi+ 2 M^3_t+M^3_l)
\cr
&\simeq V^{one-loop}_{ordinary}
-{ T  \over 12 \pi} \biggl[(M^3_\phi-m^3_\phi) + (M^3_\chi-m^3_\chi)
+ 2 (M^3_t - e^3 \phi^3) + (M^3_l- e^3 \phi^3)\biggr]
,}}
where I also used the known result
(N.B. $V^{one-loop}_{ordinary}$ is
the contribution to the effective potential from the one-loop diagram
of the ordinary loop expansion)
\eqn\ordol{
V^{one-loop}_{ordinary} \simeq
{1 \over 2} m^2 \phi^2 + {\lambda \over 4!} \phi^4
+{ T^2  \over 24} (m^2_\phi+m^2_\chi+ 3 e^2 \phi^2)
-{ T  \over 12 \pi} (m^3_\phi+m^3_\chi+ 3 e^3 \phi^3)
{}~.}
Eq.\vmlin\space shows that the
resummation of the daisy and superdaisy diagrams
amounts to a
shift (which at high temperatures is a big shift)
in the terms cubic in the masses, and that
no term linear in the effective masses is induced.

It is interesting to notice that
the daisy resummed effective
potential$^{[\kapunp,\car]}$ (in which the contributions of the superdaisy
diagrams are not included)
also amounts
to a shift of the terms cubic in the masses.
However, in that case
the squares of the effective masses are obtained by simply adding to
the squares of the tree-level masses the ordinary
one-loop self-energy contributions whereas, as shown in the preceding
sections, the daisy and superdaisy resummation requires effective
masses which are self-consistent solutions
of the gap equations \appgapm.
It is easy to verify that the $daisy-effective-masses$\space
and the $daisy+superdaisy-effective-masses$\space only agree to order $eT$.

**************************************************************************
The gap equations \appgapm\space combined with the following expression
of $V^{res}_T(\phi,G(\{M\}))$
\eqn\vresult{\eqalign{
V^{res}_T(\phi,\{ M \}) \simeq &
{1 \over 2} m^2 \phi^2 + {\lambda \over 4!} \phi^4
+{ T^2  \over 24} (m^2_\phi+m^2_\chi+ 3 e^2 \phi^2) \cr
&-{ T  \over 12 \pi} (M^3_\phi+M^3_\chi+ 2 M^3_t+M^3_l)
\cr
&+ { T M_\phi  \over 8 \pi} [M^2_\phi - m^2_\phi]_{O(T^2)}
+ { T M_\chi  \over 8 \pi} [M^2_\chi - m^2_\chi]_{O(T^2)}
\cr
&+ 2 { T M_t  \over 8 \pi} [M^2_t - e^2 \phi^2]_{O(T^2)}
+ { T M_l  \over 8 \pi} [M^2_l - e^2 \phi^2]_{O(T^2)}
\cr
& - {e^2 T^3  \over 96 \pi} (4 M_l + 3 M_\phi + 3 M_\chi)
- {\lambda T^3  \over 144 \pi} (M_\phi + M_\chi)
,}}
which has been obtained from \vtcl, \vii, \viic, and \restv,
give the desired
Rayleigh-Ritz$^{[\ref\rref{BLA BLA RAYLEIGH-RITZ.}]}$
and high-temperature approximation of the daisy and superdaisy resummed
finite temperature effective potential for the
Abelian Higgs Model.
Note that in \vresult\space
the high-temperature and small coupling expansion is truncated to order
$ T M^3$ (as it can be seen from \appgapm,
$M_{\phi,\chi,l} \sim e T$ and
$e^2 T^3 M_{\phi,\chi,l} \sim T M_{\phi,\chi,l}^3$),
and the physically
irrelevant $\phi$-independent contributions have been omitted.

**************************************************************************

$^{[\ref\rref{The denomination ``Rayleigh-Ritz approximation"
comes from many-body theory where analogous mean-field techniques have
been used since several years.}]}$
**************************************************************************
$^{[\ref\mtmldiff{Instead the effective mass $M_l$ of the longitudinal
mode of the gauge boson can be different from the effective mass $M_t$ of
the two transverse modes essentially because of the breaking of Lorentz
invariance in finite temperature field theory.}]}$,
**************************************************************************
As required for consistency of the approximation which is used in
the following (see $[\corn]$),

**************************************************************************
(which I leave to the
interested experts of these numerical techniques),

**************************************************************************
Moreover, one can observe that
in Landau gauge all the vacuum-to-vacuum diagrams
of the ordinary loop
expansion of $V_T$ which involve $D_{a \mu}$ vanish.
Therefore, in this
hypothesis (diagonal $D^{-1}_{ab}$ and Landau gauge),
all the diagrams of the ordinary loop expansion which are effectively
resummed by $V^{(a)}_{2}(G_{12})$
(the part of $V^{(a)}_{2}$ which depends on the nondiagonal
components of $G_{ab}$) and by $V^{(d)}_{2}$ give vanishing contributions
to $V_T$.
In the following I shall therefore
neglect $V^{(a)}_{2}(G_{12})$ and $V^{(d)}_{2}$;
in particular, see \hfgapahm, this leads to trivial
gap equations for $G_{12}$ and $G_{a \mu}$:
\eqn\gapnondia{G_{12}=D_{12}=0~~,~~~~G_{a \mu}=D_{a \mu}
{}~.}
**************************************************************************
I also like to thank L. Jacobs and C. Rebbi for discussions
concerning the duability of the numerical evaluation of $V_T^{res}$
as obtained in Sec.II, and
J. Kapusta for comments on the
evaluation of the diagram in Fig.2c.

**************************************************************************
However, due to combinatoric problems,
some of these
techniques$^{[\ref\imonlo{In particular, as shown
in Refs.$[\firef,\gacpi]$,
the ``improved one-loop method" (which consists of replacing
in the ordinary one-loop result
the tree-level
propagators by temperature dependent
effective propagators obtained by summing
certain classes of self-energy graphs)
performs the incorrect resummation of the
superdaisy diagrams.}]}$
do not actually perform the exact
resummation of the diagrams and this leads to incorrect results especially
concerning the next-to-leading contributions in the high temperature
expansion of $V_T(\phi)$.
Moreover, even when the combinatorics is correct, the calculations are
performed by neglecting the dependence of some subgraphs on the
external
momentum$^{[\ref\notarn{This is effectively realized by suppressing the
momentum dependence of some effective masses $M(k)$ with the assumption
that $M(k) \simeq M(0)$.
P. Arnold in $[\arn]$ already
pointed out the inadequacy of this assumption.}]}$,
and this introduces an uncontrolled error in the
evaluation of the superdaisy diagrams of gauge
theories.

**************************************************************************

In conclusion the method proposed in Ref.$[\gacpi]$ has been extremely
useful in the study of the Abelian Higgs Model.
Not only it allows
the exact resummation of the
daisy and superdaisy contributions to the effective potential
with the help of some (unfortunately non-simple) numerical evaluations,
but it also naturally leads to the Rayleigh-Ritz approximation in which
the self-consistency of the method is used to investigate the
analytic properties of the effective potential
with a mean-field approximation.

**************************************************************************
In preceding attempts of
resummation of the daisy and superdaisy contributions to the
effective potential of gauge theories
at some point
momentum dependent effective masses $M(k)$ also appear.
Again, in order to proceed analytically it is necessary to {\it eliminate}
the momentum dependence, and this is done
(not having the possibility to rely on the
self-consistency between the effective potential and the gap equations)
by making the {\it bold}
assumption that $M(k) \simeq M(0)$.
Evidently a much more satisfactory account of the momentum dependence
of $M(k)$ is obtained using the mean-field technique
described in Sec.III, and
therefore, even setting aside the eventuality of combinatorics problems,
the method proposed in ${[\gacpi]}$  (and its Rayleigh-Ritz version)
should be preferred in the study of gauge theories.

**************************************************************************
 the Eqs.\hf\space express
a self-consistent daisy
and superdaisy resummed effective potential
obtained from a variational
problem in which one
performs arbitrary variations of the functions $G(k)$.
**************************************************************************
In order to present the analysis in a detailed but still economic fashion,
the discussion is concentrated on
the case of the Abelian Higgs Model; however,
this simple theory already includes
all the most important ingredients
typical of gauge theories
which are relevant to the composite operator method.
Moreover, the phase transition
of the Abelian Higgs Model has recently received
a lot of attention$^{[\taka-\arndue]}$ (mostly as a ``toy
model" of the electroweak phase transition); therefore,
another interesting result of my analysis will be the
evaluation of the effective potential associated to this phase transition.
**************************************************************************
Unfortunately, the system of gap equations \hfgapahm\space cannot be solved
analytically even in the high-temperature approximation.
This technical difficulty is absent in the study at leading order of the
$\lambda \Phi^4$ theory,
for which the solution can be easily
found analytically$^{[\gacpi]}$ (in the high-temperature approximation)
because of the simple form taken by the gap equation once it is rewritten
using the {\it ansatz}
\eqn\anslf{G^{-1}_{\Phi}(k)= \cm^2(k) - k^2
{}~;}
\noindent
in particular, the gap equation
implies that the effective mass $\cm$ is a constant function
of the momentum $k$.
**************************************************************************
that in the composite operator method,
by exploiting the self-consistency  between the effective potential and the
gap equations, one has a more satisfactory way to deal
analytically with
the momentum dependence of the effective masses.
**************************************************************************
the study of the system of equations \hfgapahm\space no such
simplification occurs.
Let us see this explicitly by using
the following {\it ansatz} for the
propagators (which is the Abelian Higgs Model analog of the
{\it ansatz} \anslf):
**************************************************************************
**************************************************************************
**************************************************************************
**************************************************************************
**************************************************************************